\documentclass[10pt]{article}

\usepackage{basecompute}
\usepackage{array}

\renewcommand{\basecomputelogo}{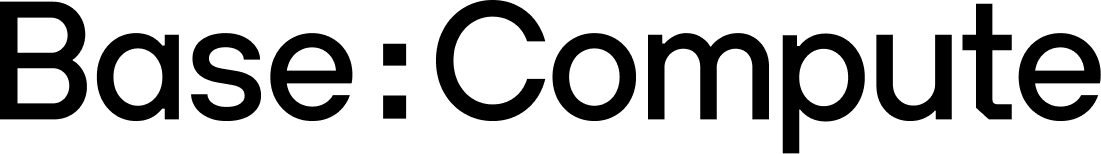}

\usetikzlibrary{arrows.meta, positioning, fit, backgrounds, calc}

\definecolor{tbdred}{HTML}{B3261E}

\basetitle{Superfluid: Local AI Server\\ for Distributed LLM Inference}

\author{
  Prabod Rathnayaka\thanks{Correspondence: \href{mailto:prabod@basecompute.co}{prabod@basecompute.co}} \quad Isura Manchanayake \quad Srikandabala Kogul \\
  Steve Siu \quad Fabian Waschkowski \quad Lukas Wesemann \\[8pt]
  \textit{Base Compute, Melbourne, Australia} \\
}

\date{}

\begin{document}

\maketitle

\begin{abstract}
Running large language models (LLMs) on local devices keeps data private, removes per-token cost, cuts network latency, and keeps working without a connection. Two shifts now strain how such models are served. On one device, several agents share a locally served model with an interactive user. Across devices, a developer or small team often owns several capable machines from different vendors, which sit idle most of the day because every local server serves one runtime on one machine. We present Superfluid, an open-source LLM server built for local inference at both scales. On one device, Superfluid drives llama.cpp, MLX, or BaseRT in isolated worker processes through one scheduler that moves interactive requests ahead of agent work, computes a prompt shared by many requests once, and writes every token to a durable log before it reaches the client. Across devices, one Superfluid server turns a fleet of machines into a single API endpoint. Machines with different hardware serve copies of the same model, receive work according to their measured speed, and can join or fail under load without losing a request. For Qwen3.8-27B on an Apple M5 Pro, a chat sent while eight agents are working receives its first token in $1.5$\,s, against $291$--$612$\,s on \texttt{llama-server}, Ollama, and \texttt{mlx\_lm.server}. On six machines from three chip vendors serving 64 simulated users for one hour, Superfluid answers all 5,041 requests while machines join one at a time and one is killed and rejoins; throughput grows from $112$ to $534$\,tok/s and the median first-token time falls from $99$ to $12$\,s. On twelve Mac Studio M3 Ultras it sustains $10{,}656$\,tok/s across 38,563 requests from 384 agents without a failure. Superfluid is available under the Apache-2.0 licence at \url{https://github.com/basecompute/superfluid}.
\end{abstract}

\section{Introduction}

Local inference has become a practical alternative to hosted APIs. A laptop with 32\,GB of unified memory serves a quantised 35B model, and runtimes such as llama.cpp~\cite{llamacpp}, MLX~\cite{mlx}, and BaseRT~\cite{basert2026} make that model fast. Running the model on the device that uses it has four advantages. Prompts, documents, and code never leave the machine (\emph{privacy}). There is no per-token bill, so a long agent run costs nothing beyond the electricity (\emph{cost}). There is no network round trip and no shared queue in a data centre (\emph{latency}). The model keeps working on a plane, behind a firewall, or during a provider outage (\emph{availability}).

Two shifts change what a local server has to do. First, local workloads are becoming \emph{agentic}. A coding agent works through a repository, another summarises mail, another reads documents, and the user still asks questions in between. This is a multi-agent workload on one device, with long and growing prompts, many concurrent requests, and one user who expects fast replies. Second, local compute is \emph{fragmented}. A developer or a small office owns several capable machines, such as MacBooks, Mac Studios, an AMD workstation, or an NVIDIA GB10, each running whichever runtime suits its hardware. Most of these machines sit idle most of the day, while the one machine serving a model saturates.

The servers that ship with local runtimes were not built for either shift. \texttt{llama-server}, \texttt{mlx\_lm.server}, and Ollama~\cite{ollama} each serve one runtime on one machine and treat every request alike. A user's question waits behind the agents' prefills, a worker crash loses the conversation in flight, and a second machine adds nothing unless the user routes requests to it by hand.

We built Superfluid for local inference at both scales. Superfluid is one server that separates the serving layer from the runtime. On one device, it treats robustness and responsiveness as first-class concerns: each model runs in its own worker process, interactive requests preempt agent work, and every token is durable before it is delivered. Across devices, the same separation lets a Superfluid head drive Superfluid nodes on other machines. Each node runs the runtime that suits its hardware, and clients see one endpoint. Our contributions are:

\begin{itemize}
  \item \textbf{A local LLM server built for agentic workloads} (Section~\ref{sec:system}): process isolation across llama.cpp, MLX, and BaseRT; a scheduler with QoS classes, preemption, and shared-prefix coalescing; and durable sessions.
  \item \textbf{Fleet mode for heterogeneous devices} (Section~\ref{sec:fleet}): machines with different hardware serve copies of one model behind one endpoint, with placement by measured speed, prefix affinity, and failover that delivers no token twice.
  \item \textbf{Evaluation at both scales} (Section~\ref{sec:eval}): on one Apple M5 Pro against each runtime's own server, and on fleets of six heterogeneous machines and twelve Apple M3 Ultras.
\end{itemize}

\section{Related Work}
\label{sec:related}

\paragraph{Data centre serving.} Orca~\cite{yu2022orca} introduced iteration-level scheduling, the basis of continuous batching. vLLM~\cite{kwon2023pagedattention} pages the KV cache, SGLang~\cite{zheng2024sglang} shares cached prefixes through a radix tree, and Sarathi-Serve~\cite{agrawal2024sarathi} chunks long prefills so that they do not stall running streams. DistServe~\cite{zhong2024distserve} places prefill and decode on different GPUs, and Helix~\cite{mei2025helix} plans model placement across heterogeneous data centre GPUs. Superfluid borrows continuous batching, prefix reuse, and chunked prefill from this line of work. It targets a different setting: one to dozens of consumer machines with different runtimes, where durability and responsiveness to one user matter more than peak throughput.

\paragraph{Local servers.} \texttt{llama-server} and \texttt{mlx\_lm.server} are the reference servers of llama.cpp and MLX, and Ollama adds model distribution around a llama.cpp server. Each is bound to one runtime on one machine, and none keeps a durable conversation record.

\paragraph{Distributed local inference.} exo~\cite{exo}, llama.cpp's RPC backend~\cite{llamacpp}, and Petals~\cite{borzunov2023petals} split one model across devices, so that a model too large for any single device can run at all. Every token then crosses the network once per stage, and one slow or departing device stalls or stops the whole pipeline. Superfluid instead replicates a model that fits on each device, places whole sessions on devices, and moves only token ids over the network. A device that is slow receives less work, and a device that leaves costs only its share of the throughput. The two approaches are complementary (Section~\ref{sec:conclusion}).

\section{Superfluid}
\label{sec:system}

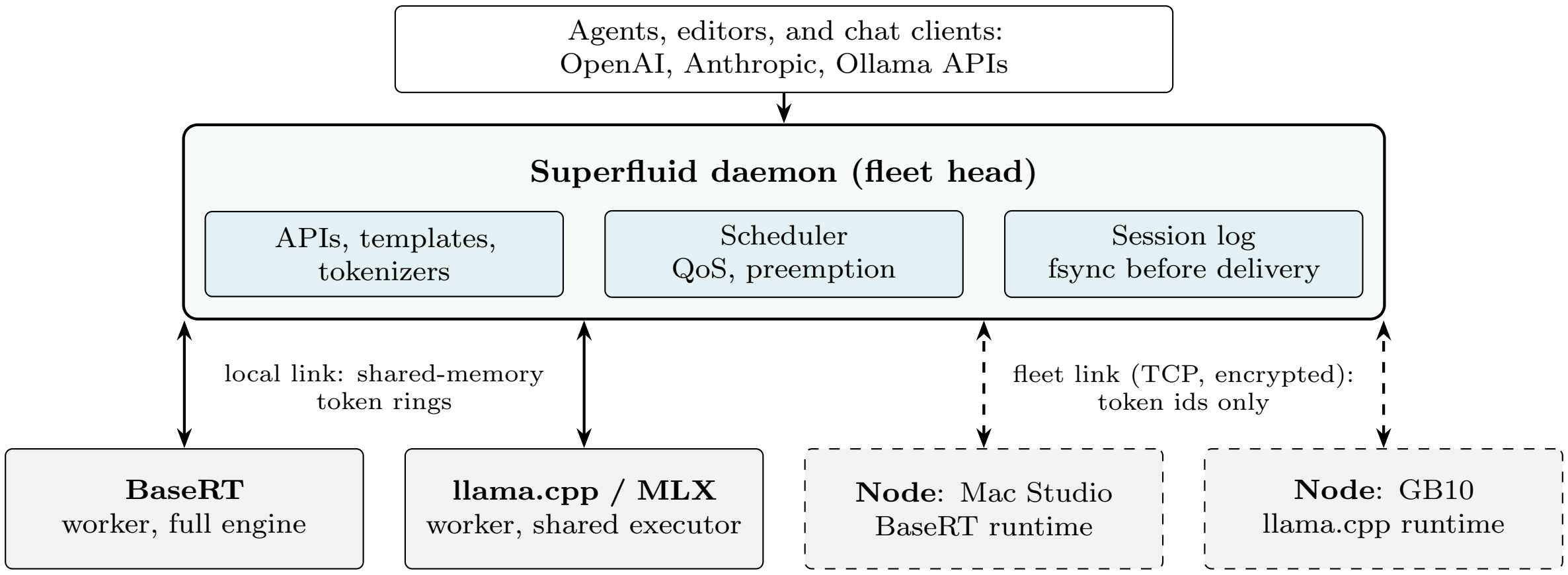
\begin{figure}[t]
  \centering
  \begin{tikzpicture}[
    font=\footnotesize,
    box/.style={draw, rounded corners=2pt, align=center, inner sep=4pt},
    comp/.style={box, fill=basecompute!12, minimum height=0.85cm, text width=3.3cm},
    wrk/.style={box, fill=black!5, minimum height=1.2cm, text width=3.3cm},
    client/.style={box, fill=white, minimum height=0.8cm, text width=7.5cm},
    arr/.style={-{Stealth[length=2mm]}, thick},
    dbl/.style={{Stealth[length=2mm]}-{Stealth[length=2mm]}, thick},
  ]
    \node[client] (c) at (0,4.8) {Agents, editors, and chat clients: \\ OpenAI, Anthropic, Ollama APIs};
    \node[font=\small\bfseries] (dt) at (0,3.55) {Superfluid daemon (fleet head)};
    \node[comp] (api)   at (-4.0,2.75) {APIs, templates,\\ tokenizers};
    \node[comp] (sch)   at (0,2.75)    {Scheduler\\ QoS, preemption};
    \node[comp] (wal)   at (4.0,2.75)  {Session log\\ fsync before delivery};
    \begin{scope}[on background layer]
      \node[draw, thick, rounded corners=4pt, fill=basecompute!4, inner sep=6pt,
            fit=(dt)(api)(sch)(wal)] (d) {};
    \end{scope}
    \node[wrk] (w1) at (-6,0.2) {\textbf{BaseRT}\\  worker, full engine};
    \node[wrk] (w2) at (-2,0.2) {\textbf{llama.cpp / MLX}\\ worker, shared executor};
    \node[wrk, dashed] (n1) at (2,0.2)  {\textbf{Node}: Mac Studio\\ BaseRT runtime};
    \node[wrk, dashed] (n2) at (6,0.2)  {\textbf{Node}: GB10\\ llama.cpp runtime};
    \draw[arr] (c.south) -- (c.south |- d.north);
    \draw[dbl] (w1.north) -- (w1.north |- d.south);
    \draw[dbl] (w2.north) -- (w2.north |- d.south);
    \draw[dbl, dashed] (n1.north) -- (n1.north |- d.south);
    \draw[dbl, dashed] (n2.north) -- (n2.north |- d.south);
    \node[font=\scriptsize, align=center] at (-4,1.4) {local link: shared-memory \\ token rings};
    \node[font=\scriptsize, align=center] at (4,1.4) {fleet link (TCP, encrypted): \\ token ids only};
  \end{tikzpicture}
  \caption{Superfluid's processes. The daemon owns the APIs, the scheduler, and the session log. Each local model runs in its own worker process; in fleet mode, nodes on other machines serve copies of the same model, each on the runtime that suits its hardware.}
  \label{fig:arch}
\end{figure}

Superfluid is written in Rust and runs as three kinds of process (Figure~\ref{fig:arch}). The \emph{daemon} owns every API, the chat templates and tokenizers, the scheduler, and the session log. One \emph{worker} per loaded model owns the weights, the KV cache, and sampling. On other machines, \emph{nodes} serve the same model for a daemon acting as fleet head.

\subsection{Architecture}
\label{sec:runtimes}

Every runtime is driven through one call: \texttt{tick(plan)}. The daemon sends a plan of admissions, prefill chunks, decode grants, and retirements. The worker sees only token ids and returns the tokens each lane produced. Daemon and worker communicate over a \emph{local link}: framed control messages travel over a Unix socket, while prompts and generated tokens travel through shared-memory rings mapped into both processes. A runtime plugs in at one of two levels. BaseRT implements the tick itself. llama.cpp and MLX implement about a dozen primitives (create, copy, truncate, and free a sequence, and run one batched step), and a shared executor builds continuous batching, prefix caching, grammar-constrained decoding, and sampling on top. A new runtime therefore inherits the whole server from one adapter crate.

Process isolation is a robustness feature. A GPU fault, a runtime bug, or an out-of-memory kill takes down one worker, not the server: the worker's streams end with a typed error, the worker is respawned, and serving continues. A request that needs a capability the model lacks, such as an image sent to a text-only runtime or a JSON schema no grammar can be compiled for, is refused with a typed error rather than served incorrectly. The OpenAI, Anthropic, and Ollama APIs all run through the same path, so tool calls, structured output, and errors behave the same on every runtime.

\subsection{Scheduling for Agentic Workloads}
\label{sec:sched}

One scheduler per model runs a continuous tick loop, and four mechanisms keep one user responsive beside many agents.

\emph{QoS classes and preemption.} Requests carry one of four classes (\texttt{interactive}, \texttt{completion}, \texttt{agent}, and \texttt{background}), set per session, per header, or per API key. When an interactive request finds every lane busy, the scheduler preempts the lowest-class lane and requeues it with its decode state. The preempted lane later replays its generated tokens as decode steps rather than as a prefill, so it continues exactly as it would have without the interruption. Waiting jobs age one class per period, so background work is never starved.

\emph{Short ticks while a request could join.} A tick is sized by duration rather than by step count, because a new request waits for the tick in flight to end. While a lane is free or an interactive request is queued, a tick carries at most about $0.5$\,s of prefill work; a lone long prompt that nothing could join runs in long ticks that use the GPU efficiently. This is chunked prefill~\cite{agrawal2024sarathi}, applied only when it helps.

\emph{Shared-prefix hold.} A prefix enters the cache only when its lane retires, so a burst of requests sharing one long system prompt would each compute it in parallel. Admission instead holds a request back briefly when a running lane shares at least 1024 of its prompt tokens and is about to finish, then seeds it from the cached prefix.

\emph{Queueing instead of failing.} When the KV cache cannot hold a plan, the engine rejects it without changing state, and the scheduler retries with half the admissions, requeuing the rest. A client that disconnects, streaming or not, frees its lane at the next tick.

\begin{table}[t]
\centering
\caption{Local serving on Qwen3.8-27B at 4 bits, Apple M5 Pro (64\,GB). Each Superfluid runtime is compared with the servers for the same model format. Arrows give the better direction; the best value within each format is in bold. A dash marks a scenario the server cannot run (it has no separate worker process).}
\label{tab:local}
\footnotesize
\setlength{\tabcolsep}{3.5pt}
\renewcommand{\arraystretch}{1.2}
\begin{tabular}{@{}>{\raggedright\arraybackslash}p{0.30\linewidth} >{\raggedright\arraybackslash}p{0.19\linewidth} rrr rr@{}}
\toprule
 & & \multicolumn{3}{c}{GGUF (llama.cpp)} & \multicolumn{2}{c}{MLX} \\
\cmidrule(lr){3-5}\cmidrule(l){6-7}
Scenario & Measure & Superfluid & \texttt{llama-server} & Ollama & Superfluid & \texttt{mlx\_lm.server} \\
\midrule
Chat sent behind 8 busy agents & first token (s) $\downarrow$ & \textbf{1.47} & 291 & 380 & \textbf{2.01} & 612 \\
32 requests share one 4k-token prompt & from cache (\%) $\uparrow$ & \textbf{97} & 75 & \textbf{97} & \textbf{69} & 0 \\
 & all served (s) $\downarrow$ & \textbf{56} & 171 & 88 & \textbf{418} & 907 \\
KV cache oversubscribed $4\times$ & completed, of 32 $\uparrow$ & \textbf{32} & \textbf{32} & \textbf{32} & \textbf{32} & 24 \\
Inference worker killed mid-stream & back (s) & \textbf{1.75} & -- & 1.83 & \textbf{1.85} & -- \\
Non-streaming clients disconnect & lanes freed (s) $\downarrow$ & \textbf{1.01} & 1.74 & 7.25 & \textbf{2.67} & 839 \\
\bottomrule
\end{tabular}
\end{table}

\subsection{Durable Sessions}
\label{sec:sessions}

A session is an append-only log of the exact token ids that each turn put on the model's input. Every record is written and \texttt{fsync}ed before the client sees its tokens, following the write-ahead discipline of database recovery~\cite{mohan1992aries}, and engine state (KV cache, prefix cache) is only a cache that can be rebuilt from the log. After a worker crash or a server restart, a session continues by recomputing its recorded prompt. Sampling uses a counter-based random generator indexed by stream position, so a seeded session draws the same tokens whether it was interrupted, resumed, or forked. A session API can continue, fork, or edit a conversation without copying its history.

\subsection{Fleet Mode for Distributed Devices}
\label{sec:fleet}

In fleet mode, one Superfluid daemon runs as a \emph{head} that owns the API, chat template, and tokenizer, and other machines join it as \emph{nodes} with \texttt{superfluid node join}. Nodes are found on the local network by mDNS or given by address, and each node installs and runs the runtime that suits its own hardware (for example, BaseRT on a Mac and llama.cpp on a CUDA workstation). Every node serves a full copy of the same model, and the head refuses a node whose tokenizer differs from its own, so that a session's token ids mean the same thing on every node.

\emph{Sessions cross the network.} The head places each session as a whole on one node, which batches it with its other sessions, samples its tokens, and streams them back. Only prompt token ids and sampled token ids cross the network; weights, KV cache, and activations never do. The fleet link carries the same frames as the local link over TCP. It is secured with a Noise protocol handshake~\cite{perrin2018noise} keyed by a fleet token that the head generates: only a machine holding the token can join, and every frame is then encrypted and authenticated.

\emph{Placement by measured speed.} Nodes report their measured prefill and decode rates, active lanes, queue depth, and KV occupancy. For each new session, the head filters out nodes whose context window is too small or whose KV cache is nearly full, and estimates on each remaining node the time to finish the work already queued ahead plus the new request's own prefill and decode. It then picks the node with the earliest estimated finish. A fast GB10 therefore receives more sessions than an older laptop, without any configuration. A conversation's next turn is placed first on the node that holds its longest cached prefix, so a multi-turn agent does not recompute its history on a different machine.

\emph{Joining, leaving, and failing.} A node that joins receives new sessions immediately. A node that leaves or fails mid-stream is marked unhealthy; the tokens it already delivered become part of the prompt, and the session is placed on the next capable node, which generates the rest. The client sees one uninterrupted stream in which no token is delivered twice.

\section{Evaluation}
\label{sec:eval}

We evaluate Superfluid as a local server on one device (Section~\ref{sec:eval-local}) and as a fleet across heterogeneous devices (Section~\ref{sec:eval-fleet}). Every number is derived from per-request logs written by the load generators in the repository~\cite{superfluid}.

\subsection{Local Server}
\label{sec:eval-local}

\begin{figure}[t]
  \centering
  \includegraphics[width=\linewidth]{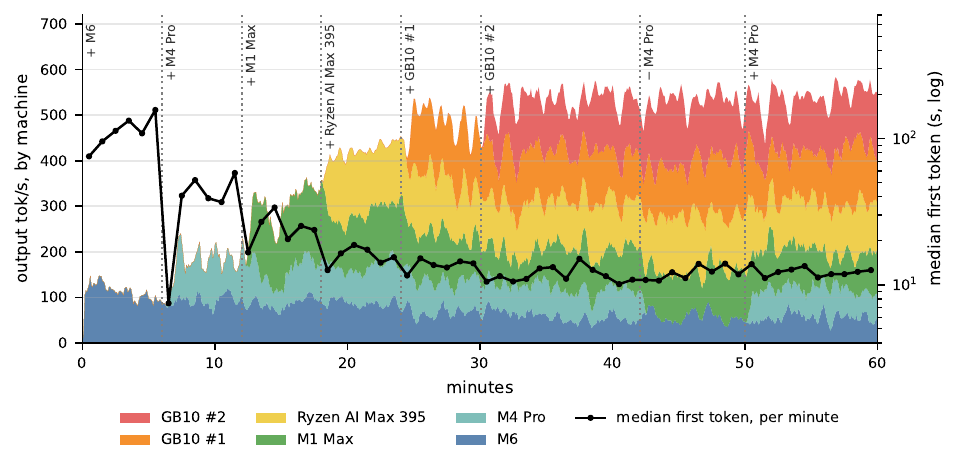}
  \caption{Run A: six heterogeneous machines as one endpoint, Qwen3-4B on llama.cpp, 64 simulated users for one hour. Stacked areas give the output tokens per second each machine served (30\,s rolling rate); the line gives the median time to first token per minute (right axis, log scale). Dotted lines mark machines joining ($+$) and the node killed at 42 minutes ($-$). All 5,041 requests were answered.}
  \label{fig:fleet}
\end{figure}

\paragraph{Setup.} We run Qwen3.8-27B, a hybrid of linear-attention and full-attention layers, at 4 bits on an Apple M5 Pro with 64\,GB: the Q4\_K\_M GGUF under llama.cpp (build b11284), compared with \texttt{llama-server} and Ollama v0.35.1, and a 4-bit MLX conversion under MLX (mlx-lm 0.31.3), compared with \texttt{mlx\_lm.server}. Each Superfluid runtime uses the same runtime build as the server it is compared with, and every system has the same number of lanes and the same context per lane. Each scenario starts every system fresh and uses greedy decoding with thinking off.

\paragraph{Results.} Table~\ref{tab:local} summarises the results. The largest difference is in priority under load: a chat sent while eight agents prefill receives its first token in $1.5$\,s on llama.cpp and $2.0$\,s on MLX through Superfluid, while the other servers queue it behind the agents for $291$--$612$\,s. When 32 requests share one 4k-token system prompt, the shared-prefix hold computes the prompt about once for the whole burst and serves it in $56$\,s, against $171$\,s on \texttt{llama-server}, which recomputes it once per lane. With the KV cache oversubscribed four times, every request completes on Superfluid, while \texttt{mlx\_lm.server} loses 8 of 32 to timeouts and reset connections. A killed worker on Superfluid ends its stream with a typed error, and the server serves again in under 2\,s, where Ollama truncates the stream and the single-process servers go down with the worker.

This costs nothing in raw speed. On Qwen3-4B at 4 bits on an Apple M1 Max, Superfluid on llama.cpp matches \texttt{llama-server} at one request ($76.7$ against $76.8$\,tok/s) and is $4\%$ ahead at eight ($150.8$ against $145.1$\,tok/s). Because \texttt{mlx\_lm.server} serves one request at a time, Superfluid's shared executor turns MLX into a batching server with $1.35\times$ its aggregate throughput at eight requests ($130.2$ against $96.8$\,tok/s).

\subsection{Fleet Mode}
\label{sec:eval-fleet}

\paragraph{Office workload.} Our main fleet experiment simulates an office. Sixty-four users hold multi-turn conversations for one hour, each drawing a task (a question, an email, a 2,400-word document, or a coding task with 2--5 turns), waiting for the streamed reply, and pausing for varying periods of time (mean 6\,s) before the next turn. The workload is closed-loop, so a faster fleet receives more requests. A GB10 acts as the head, and every node runs 8 lanes with 16k tokens of context. Machines join one at a time every 6 minutes; at 42 minutes we kill the Superfluid node on one machine mid-stream and restart it 8 minutes later. We run the workload twice:
\begin{itemize}
  \item \textbf{Run A}: Qwen3-4B-Instruct-2507 (Q4\_K\_M) on llama.cpp, on six machines from three chip vendors: an Apple M6, an M4 Pro, an M1 Max, an AMD Ryzen AI Max 395, and two NVIDIA GB10 systems.
  \item \textbf{Run B}: Qwen3.6-35B-A3B (4-bit) on BaseRT, on Metal and CUDA: an Apple M6, an M4 Pro, an M1 Max, an M5 Pro, an M3 Ultra, and a GB10.
\end{itemize}

\begin{table}[t]
\centering
\caption{Office workload by fleet size. Output throughput comes from the nodes' decode counters, and first-token times from the requests started in each interval. The row with six machines covers minutes 30--42, before the node is killed.}
\label{tab:fleet}
\footnotesize
\setlength{\tabcolsep}{5pt}
\begin{tabular}{@{}l rrr rrr@{}}
\toprule
 & \multicolumn{3}{c}{Run A: Qwen3-4B, llama.cpp} & \multicolumn{3}{c}{Run B: Qwen3.6-35B-A3B, BaseRT} \\
\cmidrule(lr){2-4}\cmidrule(l){5-7}
Machines & output tok/s & first token p50 (s) & p90 (s) & output tok/s & first token p50 (s) & p90 (s) \\
\midrule
1 & 112 & 98.5 & 309.1 & 104 & 66.3 & 198.8 \\
2 & 180 & 42.5 & 147.9 & 155 & 41.7 & 107.8 \\
3 & 304 & 24.8 & 70.8 & 214 & 26.8 & 73.4 \\
4 & 424 & 16.1 & 30.0 & 391 & 9.5 & 23.6 \\
5 (joining) & 481 & 13.8 & 24.1 & 469 & 7.7 & 16.1 \\
6 & 534 & 11.7 & 20.7 & 476 & 6.9 & 13.4 \\
5 (one killed) & 532 & 12.0 & 21.5 & 444 & 7.7 & 18.5 \\
\midrule
Whole hour & 417 & 13.4 & 30.8 & 367 & 8.1 & 28.1 \\
Requests (failed) & \multicolumn{3}{c}{5,041 (0)} & \multicolumn{3}{c}{7,151 (0)} \\
\bottomrule
\end{tabular}
\end{table}

\paragraph{Results.} Figure~\ref{fig:fleet} and Table~\ref{tab:fleet} show the fleet absorbing load as machines join. In Run A, a single M6 is overwhelmed by 64 users (median first token $99$\,s); each machine that joins takes a share of the load at once, and with six machines throughput reaches $534$\,tok/s ($4.8\times$) and the median first token falls to $12$\,s. Run B, a larger mixture-of-experts model on a different runtime, behaves the same way: throughput rises from $104$ to $476$\,tok/s and the median first token falls from $66$ to $7$\,s. The machines contribute according to their speed: in Run A, the two GB10s and the Ryzen AI Max each served $114$--$122$\,tok/s while they were up, against $71$--$73$\,tok/s for the M4 Pro and the M6, without any per-machine configuration.

Killing a node mid-stream costs no requests. In Run A, 61 requests were in flight across the fleet when the M4 Pro's node was killed, and all completed: the head moved each stream on the killed node to another machine, which continued it from the tokens already delivered. In Run B, all 52 requests in flight completed. Across both runs, all 12,192 requests were answered. Throughput with five machines after the kill matches throughput with six in Run A ($532$ against $534$\,tok/s), which suggests that from about five machines on the fleet is limited by the 64 users' demand rather than by its capacity.

\paragraph{Scaling throughput.} To measure capacity rather than demand, we ran a closed-loop agent sweep on twelve Mac Studio M3 Ultras (512\,GB), with Qwen3.6-35B-A3B on BaseRT 0.4.6, 128 lanes per node, and no speculative decoding. Each agent sends its next request as soon as its last reply ends, every prompt is unique, and every reply has 256 tokens. One machine also acted as the head and ran the load generator. Six load points of 12 minutes each served 38,563 requests without a failure. With 512-token prompts and 192 agents, the fleet produced $3{,}072$ output tok/s at a median first token of $2.0$\,s and $20$\,tok/s per agent. With 2048-token prompts and 384 agents, it processed $10{,}656$\,tok/s in total ($9{,}249$ prompt and $1{,}407$ output).

\section{Conclusion}
\label{sec:conclusion}

Local inference is moving from one chat on one machine to many agents on many machines. Superfluid serves both: on one device, it keeps the user responsive beside eight busy agents ($1.5$\,s against $291$--$612$\,s), ends a crashed worker's streams cleanly, and keeps every conversation in a durable log. Across devices, Superfluid turns idle machines from different vendors into one endpoint that grows as they join and loses no request when one fails. Superfluid and its benchmark tools are available under the Apache-2.0 licence at \url{https://github.com/basecompute/superfluid}.

\paragraph{Limitations.} Fleet mode replicates the model, so every node must hold a full copy. A session that moves to another node recomputes its prompt, because nodes do not exchange KV caches. Nodes may differ in quantisation, in which case a session's output depends on the node that serves it. Our fleet experiments ran one runtime per fleet, and we did not compare placement policies directly. Images, audio, and speculative decoding are served only by BaseRT, and the server speaks no TLS, so it is meant for loopback, a trusted network, or a reverse proxy.

\section*{Acknowledgements}

We thank Mount Thor for early access to their Apple silicon cluster. Running on twelve of their Mac Studio M3 Ultras allowed us to evaluate Superfluid at a scale beyond our own hardware.

\bibliographystyle{unsrtnat}
\bibliography{references}

\end{document}